\documentclass{sig-alternate}
\toappear{} 
\begin{document}
%
\conferenceinfo{FDG}{'12 Chania, Crete, Greece}

\title{
Analysis of  player's in-game performance vs rating:\\
Case study of Heroes of Newerth}
%
%
%
%
%
\def\sharedaffiliation{%
\end{tabular}
\begin{tabular}{c}
}
\numberofauthors{3} 
    \author{
      \alignauthor Neven Caplar\\     
    \email{caplarn@phys.ethz.ch}
     \affaddr{Institute for Astronomy - ETH }\\
       \affaddr{Wolfgang-Pauli-Strasse 27, 8093 Zurich, Switzerland}\\
      \alignauthor Mirko Suznjevic\\      
     \email{mirko.suznjevic@fer.hr}
      \affaddr{University of Zagreb Faculty of Electrical Engineering and Computing}\\
       \affaddr{Unska 3, 10000 Zagreb, Croatia}\\
%
\alignauthor Maja Matijasevic\\    
     \email{maja.matijasevic@fer.hr}\\
     \affaddr{University of Zagreb Faculty of Electrical Engineering and Computing}\\
       \affaddr{Unska 3, 10000 Zagreb, Croatia}\\
%
          }

\maketitle
\begin{abstract}
We evaluate the rating system of ``Heroes of Newerth'' (HoN), a multiplayer online action role-playing game, by using statistical analysis and comparison of a player's in-game performance metrics and the player rating assigned by the rating system. The datasets for the analysis have been extracted from the web sites that record the players' ratings and a number of empirical metrics. Results suggest that the HoN's Matchmaking rating algorithm, while generally capturing the skill level of the player well, also has weaknesses, which have been exploited by players to achieve a higher placement on the ranking ladder than deserved by actual skill. In addition, we also illustrate the effects of the choice of the business model (from pay-to-play to free-to-play) on player population.  
\end{abstract}
\maketitle

\section{Introduction}

Aside from their huge entertainment and media business prospects, multiplayer online role-playing games (RPGs), with millions of players worldwide, are also an exciting area of research. In this paper we focus on the player rating system, using ``Heroes of Newerth'' \cite{HonWebSite} as a case study.

As the main goal of each player in an action-RPG is to advance on the player ranking ladder, the player rating system is one of the key elements for the mid- to long-term success of the game in a growing and very competitive video games market. The user experience, i.e., the player, is king, and thus the player's rating should, ideally, represent an ``objective'' measure of the player's skill. (Perceived) accuracy matters, as well as speed.
A good rating system has to place the player quickly on his/her skill level in order not to discourage the player by setting him/her against too easy, or too difficult, opponents. Also, as in all team matches, in action-RPGs it is necessary to create teams from players having comparable skill levels. If the team is unbalanced, the players are often very negative towards less skilled team-mates, creating a very hostile atmosphere for the new, or just less skilled players. 

So, on one hand, the player rating system should be (and typically is) designed so as to take into account the player's in-game performance to create the player ranking list; and the players know which actions and achievements count for the score. Most player rankings are based on the end-of-the-match score. On the other hand, knowledge of the game inner mechanisms in general, and the ranking algorithm in particular, is tempting. Taking advantage of design flaws, software bugs, as well as the regular game features in a way not intended by the game designers, has also always been ``a part of the game''. Hence, the purpose of this paper is twofold: first, it studies the relationship between player ranking and in-game performance metrics; and second, it reveals the player behaviors that attempt to exploit the weaknesses discovered in the ranking system. To the best of our knowledge there is currently no study which directly analyses the correlation of in-game player performance metrics with the rating assigned by the system, though a similar evaluation has been performed for one of the chess rating systems \cite{Fatta:SkillRating}. The results of this study may be useful for developing new matchmaking algorithms, based on a broad set of player skills and metrics to create more fun and balanced matches \cite{Delalleau:GhostRecon,Zhang:ContextSkill}.

The game used as a case study is ``Heroes of Newerth'' (HoN), a multiplayer action-RPG, developed and published by the S2 Games. We perform a statistical analysis of the performance of HoN's rating system through comparison with several in-game performance metrics. The dataset used for the analysis contains the player performance data (both the players' ratings and the metrics) for 338,681 players, but majority of the analysis is performed on a subset of the data due to computational and visual reasons. 


The remainder of the paper is organized as follows: in Section 2 we present the related work, followed by a detailed description of the HoN's rating system and game mechanics in Section 3, to make the reader familiar with the terms which will later be discussed. Section 4 contains the measurement methodology, and in Section 5 we present the results related to the correlation of players' in-game performance metrics and assigned ratings. We conclude the paper in Section 6.
\pagebreak
\section{Related work}
Various algorithms have been used for player ranking and rating in online games, from simple position swapping of the players based on the results of their match, to very complex statistics based systems. The rating system based on fuzzy logic proposed in \cite{Graf:Matchmaking} matches the players according to their estimated skill and the desired skill of the opponent. 
Some rating systems are based on overall player performance in the game \cite{Shim:Inferring}, or on player's reputation \cite{Kaiser:PlayerRating}, but in this work we focus on those intended for match-based games. An analysis comprising several datasets of match based games has been presented in \cite{Guo:MatchAnalysis}.
\\
Elo rating system \cite{Elo:Chess}, named after its creator Arpad Elo, developed for the purpose of rating chess players, has been widely used in online game rating systems, as well. A detailed analysis of Elo system for chess rating has been presented in  \cite{Glickman:Guide,Glickman:Rating}, while recent modifications for its use in USA Chess Federation have been described in \cite{Glickman:USCF}.
\\
The original Elo system has been slightly modified for different game genres which have ``match based'' structure similar to chess. For example, World of Warcraft (WoW), as the world's most popular Massively Multiplayer Online Role-Playing Game (MMORPG), has initially used a slight modification of the Elo system \cite{WoWWiki:ArenaSystem} for  match based player vs player combat (also called arena matches). The system underwent certain changes in order to fight the observed exploits (very skilled players leaving teams and fighting inexperienced players, win trading between arena teams, etc.). 
\\
The first Bayesian rating system was ``Glicko'' by Glickman \cite{Glickman:Glicko}, which estimates players' skill over a certain period. 
Its limitation is that it is designed to address only matches between two people and many online games have combat between teams of multiple players. Therefore, Microsoft has developed a ``TrueSkill'' rating systems, a Bayesian rating system which calculates the rating after every match and is able to do it for multiple players \cite{Herbrich:TrueSkill}. The TrueSkill system is designed to track the uncertainties about player skills, also models draws, and it is able to extract the individual player skills from team results. The next step in the evolution of the TrueSkill system is presented in \cite{Dangauthier:TrueSkillTime}, with addition of a smoothing system through time instead of filtering. A Bayesian approximation method for derivation of simple analytic rules for updating team strength in multi-team games has been presented in \cite{Weng:OnlineRankin}. The authors claim that the ranking results are comparable with TrueSkill, while the computation time and code are shorter.
\\
In \cite{Zhang:ContextSkill} authors propose Factor-Based Context-Sensitive Skill Rating System, which extends the approach of TrueSkill through explicitly modelling different skills based on context (e.g., in chess playing as White would be modeled differently from playing as Black). A neural networks based approach to matchmaking through evaluation of multiple skills of the player and maximizing the perceived fun has been introduced in \cite{Delalleau:GhostRecon}. Another use of neural network enables the prediction of the outcome of complex, uneven two-team group competitions by rating individuals through re-parametrization of the Bradley-Terry model \cite{Menke:NeuralRating}.
\\
Matchmaking algorithms do not have to be based only on the skill of the player. For example, matchmaking in wireless networks may be based on network characteristics of the participating players in order to create a match with highest feasible game performance quality \cite{Manweiler:SwitchBoard}.
%

\section{{HoN} mechanics and rating system}
\label{sec:mmr}

HoN was inspired by the Defense of the Ancients (DotA) map, which originated as a custom map for the Blizzard Entertainment's real time strategy game ``Warcraft III''. The game was officially released on May 12, 2010, as a pay-to-play game, and re-released as a free-to-play game on July 29, 2011. According to S2 Games, in July 2011, HoN already had 460,000 unique active players \cite{S2GamesWebSite}. 

In HoN a player controls one character, dubbed ``hero", with unique skills. Players are joined in groups of five, and assigned to one of the two combating factions, ``Hellbourne'' and ``Legion''. Each faction has a ``base'', located at the opposite corners of the map. The goal of the game is to destroy the building in the center of the enemy base; achieving this requires the players to destroy a number of enemy buildings beforehand. Each faction is also helped by its own non-player characters, called ``creeps''. Creeps have simple artificial intelligence with well defined and known rules. Creeps of the opposing faction can be destroyed for gold and experience, same as the ``neutral'' creeps, which are positioned across the map at certain locations, and do not fight for any of two factions. 

Gold and experience are resources used for improving one's in-game character. Apart from killing creeps, they can be earned by destroying buildings and, most importantly, by killing enemy heroes. When killing an enemy hero, the slaying hero is richly awarded by getting a large gold and experience bonus. The hero who has been killed has to wait for a certain amount of time before returning to the game, by ``respawning'' at the base. While ``killed'', a hero is unable to earn gold and experience for himself, and can not oppose the enemy team in achieving strategic advantage. 

Within the game, each player is assigned an individual rating.
As the player rating is applied to automatically group together players with similar skill level by a matchmaking algorithm, it is called Match Making Rating (MMR).
The default initial rating assigned to new players is 1500. 
Players typically get $\pm$ 5 points added to MMR after each match, depending on the outcome (+5 for wining, and -5 for losing). This number can slightly vary depending on the rating discrepancy between two teams and the player's rating in relation to the other players in the match. The lowest MMR set by the algorithm is 1000, so no player can drop below that rating. \\
The algorithm employed by the game is a version of Elo rating system. As previously stated, Elo rating system has been widely employed and used, but suffers from certain problems, such as rating inflation, and freezing of top rankings (by players who stop playing once they have reached top positions, i.e., no rating deflation over time) \cite{Regan:Inflation}.

The HoN's Matchmaking algorithm is also not an exception when it comes to exploits.
A notable event in its evolution was the release of patch 2.5 in December 2011, in order to ``push'' the rating closer to the initial value and assign the true skill rating faster to new players. It also addressed a recognized problem of ``smurfs''. Smurfs are very experienced players, who create new accounts (posing as entry-level players) and then play against real ``newbies'', thus winning easily, but ruining the playing experience for inexperienced, and often new, players in the process (and  cutting into the future profit for the company, as well).

\section{Methodology}
We now describe the player rating dataset and in-game performance metrics.
\subsection{Player rating dataset}
To obtain the player rating data, we have retrieved the whole player ladder (known as ``Ranked MMR Ladder (Matchmaking)''), which is made available from the HoN's web-site by its developers. We also supplemented this data with the data made available by user sites that query the game database, namely, ``HoNEdge" at www.honedge.com. By combining the data, we have obtained the statistics of 338,681 players. The data only includes players that have been deemed ``active'', i.e., those who have played more than 10 matches altogether and who have logged in into the game within the last 30 days from the date when the data was retrieved (timestamp: 19 Oct. 2012 at 23:59:59 EST).
\\
The distribution of the players according to their MMR is very well approximated by the normal curve (Figure~\ref{fig:MD}), with the mean ($\mu$) and variance ($\sigma$) being $\mu = 1528$ and $\sigma=112$. Bin sizes in Figure~\ref{fig:MD} are 10 ranking points wide, with 105 bins produced. Median of the curve is at 1527. 
There is a slight excess of the players around 1500 mark, corresponding to MMR value of 1500 for new accounts. Additionally, there are very small excesses of players at 1600, 1700, 1800, and 1900 points, which denotes the players who are not playing further games, after reaching a new threshold. Also, it is interesting to note that the mean value is higher then the default initial MMR value (1500). This ``inflation'' of points (a known phenomenon in rating systems \cite{Fatta:SkillRating}) is mainly caused by new, inexperienced players, who often lose interest in the game after some time, or, create new accounts, by that way introducing an excess of ranking points in the system. 
\\
\begin{equation}
\mathcal{P}(x)= x^{\alpha -1}(1-x)^{\beta-1}
\end{equation}
where $x$ is defined to be in range $0<x<1$. 
\begin{figure}
\centering
  \includegraphics[width=0.48\textwidth]{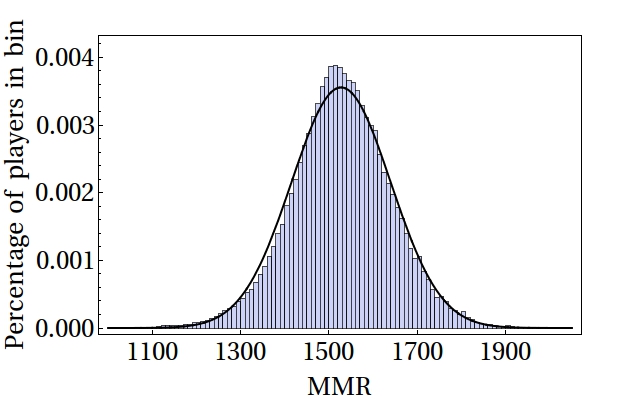}
	\caption{MMR distribution of players} 
	\label{fig:MD}
	 \end{figure} 
Recognizing that computation with such a large number of points would be difficult and taking in account that our primary interest is in how behavioural habits of players change across the range of MMR values, we construct a representative sample of $\gtrsim $ 3000 players that are approximately uniformly distributed according to MMR. The distribution of the sample can be seen in Figure~\ref{fig:PD}. Bin sizes are 50 ranking points wide, with 21 bins produced, as shown in the Figure. Construction of a representative sample has been done by modifying random seed to have higher probability to select players from a MMR range where there are fewer players. Probability distribution which was used for random seed was defined by beta distribution
Particular values for parameters $\alpha$ and $\beta$ that were used in this work were $\alpha=\beta=0.12$. This function is symmetric with a strong growth for $x$ close to zero and $x$ close to 1. After a random number is produced in that way, we multiply it by the size of the sample (338,681) and find the closest integer value. This number is matched with the player's position on the ladder, which is then added into the sample. If the value is repeated, it is discarded, since a single player can appear in the sample just once. We repeat this process until we reach the desired sample size consisting of unique players. 
It is clearly not possible to produce a large sample that would be uniform across the whole MMR range, as there are not enough players at the edges of the range. For instance, there are only 14 players in $\ge$ 2000 range, so it is not possible to find enough data points in this area. For that reason, when drawing the fitting curves, we stay in the ``safe'' MMR range between 1050 and 1950.
\begin{figure}
\centering
  \includegraphics[width=0.48\textwidth, clip, trim=0mm 0mm 0mm 9mm]{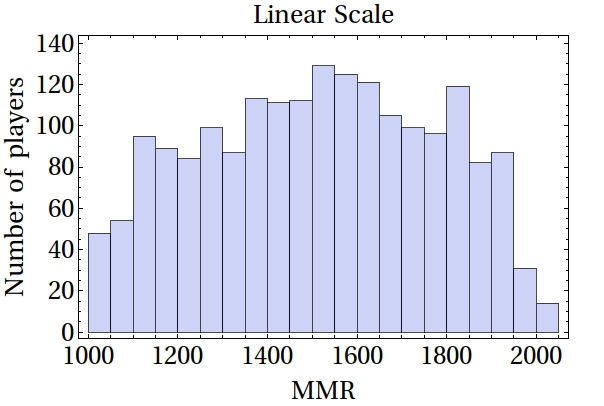}
	\caption{Number of players according to their MMR in one of produced samples} 
	\label{fig:PD}
	 \end{figure}    


\subsection{In-game performance metrics and MMR}
We studied the following in-game performance metrics: 
\begin{itemize} \itemsep1pt \parskip0pt \parsep0pt
\item number of games played, 
\item win/loss ratio, 
\item kill/death and assist/death ratio, 
\item gold per minute and experience per minute, 
\item game duration, 
\item action rate, 
\item wards per minute, 
\item denying, and lastly, 
\item account age. 
\end{itemize}
We select these metrics as they reflect the players' behaviour in the game, and also indicate the level of players' skill. By studying the correlations of these variables with the rating of the players, we reach several conclusions regarding the relationship of the players' in-game performance and rating. 
\\
To achieve this goal, we create a representative sample, as described in previous section, and note the MMR of the player and metric that is of interest. After that, we also make a fit to
\begin{equation}
1+a \cdot \mbox{MMR}+b \cdot \mbox{MMR}^{2}
\end{equation}
dependence, where $a$ and $b$ are constants. Even though it might not be best choice of fitting formula for all metrics, it has been used for consistency and ease of comparison between different metrics.

\section{Results}
We now explore the correlation between the different in-game performance metrics and MMR of the players. 
\subsection{Number of games played}
One would expect that there would be a strong correlation between the number of games played and the MMR of the player. As seen in Figure~\ref{fig:GP}, although there is correlation, there are quite a few anomalies. The scale on figure~\ref{fig:GP} is logarithmic and the best fit is shown as a black line. Firstly, it may be noticed that the lowest ranked players have relatively large number of matches played. This is a consequence of the new accounts starting at MMR of 1500  points, so a player has to lose a relatively large number of games to fall to the bottom of the ladder. 
\\
\subsection{Win/loss ratio}
\begin{figure}
\centering
  \includegraphics[width=0.48\textwidth, clip, trim=0mm 0mm 0mm 1mm]{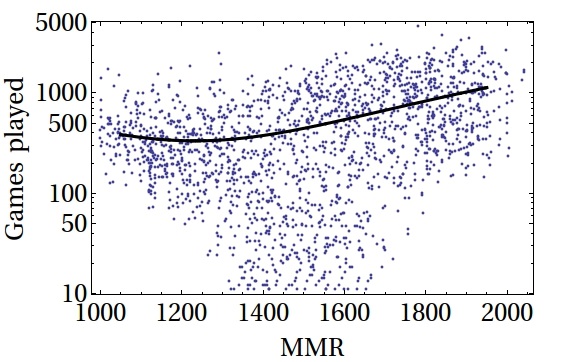}
	\caption{Number of games played as function of MMR} 
	\label{fig:GP}
	 \end{figure}
Secondly, we observe number of players that have high a MMR, although they having a far fewer games played then other players in their bracket. This can be explained by the ability to create a new account for free, and to have a paid reset of account statistics. Very good players who exploit these options are thus artificially placed in the wrong bracket, and so tend to win many consecutive games, resulting in a fast advancement up the MMR ladder.  This phenomena can also be attributed to different approaches to the game by different groups of players. While many players play ``just for fun'', (i.e., play with little desire to significantly improve their rating), some players try to improve their rating not by only playing the game, but also through other means (e.g., by watching the games played by the very best players, studying replays of matches, learning from guides available and memorizing various subtle rules that are incorporated in the game) so they can achieve higher ratings in a fewer games played. 
Effect of very good players playing below their level is even more obvious if we consider win/loss ratio of the players. We define, for each player, win/loss ratio simply as 
\begin{equation}
\mbox{win/loss ratio}= \frac{\mbox{number of wins} }{\mbox{number of losses}}.
\end{equation}
We notice that the largest number of players has win/loss ratio of 1 (Figure~\ref{fig:WR}), meaning that they are in the bracket where they truly belong, winning and losing approximately the same number of matches.
Also, there are players with low win/loss ratio in the lower range, meaning that they have to fall some more for their skill level to be accurately described by their MMR. On the other edge of the spectrum there are the accounts with very high win/loss ratio, which can be attributed to new accounts and statistics resets, as described in the previous paragraph. As one would expect, players on the edges of a MMR range have win/loss ratio substantially different from 1, meaning that they can not be matched against players of equal skill level, so they tend to win/lose substantially more. It is interesting to notice a few points that represent players who are below MMR of 1500, but have win/loss ratio significantly higher then 1. This is a result of the abuse of the system by players who play in carefully constructed groups (with help of their friends) which allows them do manipulate their rating in such a way that they gain a small number of points when winning, and lose a large amount of points when losing. In this way, they manage to have high win/loss ratio, while still having a low overall rating. This weakness of the matchmaking algorithms results in matching people with different skills in the game which degrades the experience of the lower skilled players. 
\begin{figure}
\centering
  \includegraphics[width=0.48\textwidth, clip, trim=0mm 0mm 0mm 10.5mm]{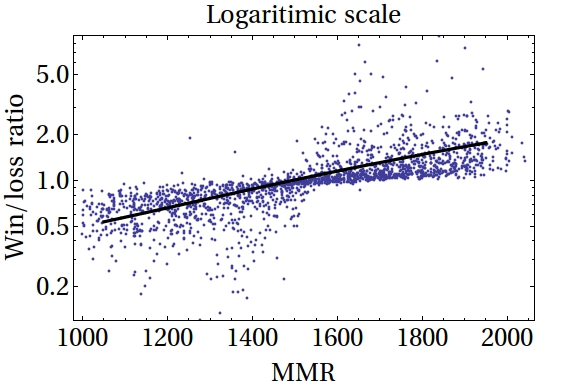}
	\caption{Win/loss ratio as function of MMR} 
	\label{fig:WR}
	 \end{figure}  

 \subsection{Kill/death and assist/death ratios}	 
	 
As mentioned earlier, the main mechanism to take advantage in a game is killing enemy heroes. A ``kill'' is granted to the player that lands the killing blow on the enemy unit (hero or creep). It is also possible that a hero does not get the credit for the kill, because, for example, the enemy hero has been killed by neutral creeps or by his own teammates (in special and rare circumstances). This results in overall kill/death ratio in our sample to be 0.957. We define, for each player, kill/death ratio, K/D, as 
\begin{equation}
\mbox{K/D}= \frac{\mbox{number of kills}}{\mbox{number of deaths}}.
\end{equation}

Additional mechanics in the game is assisting. An ``assist'' is granted to a hero that deals a damage to the enemy hero, who is then killed within the next 18 seconds, but does not land the killing blow. Assisting in a kill also brings gold to the hero in question, although in a significantly smaller amount then to the killing hero. \\
\begin{figure}
\centering
  \includegraphics[width=0.48\textwidth, clip, trim=0mm 0mm 0mm 9mm]{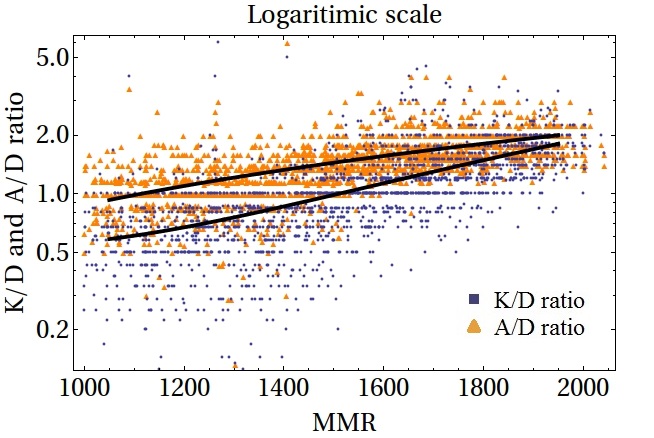}
	\caption{K/D and A/D as a function of MMR} 
	\label{fig:KDA}
	 \end{figure} 
We define, for each player, assist/death ratio, A/D, as 
\begin{equation}
\mbox{A/D}= \frac{\mbox{number of assists}}{ \mbox{number of deaths}}.
\end{equation}
Figure \ref{fig:KDA} shows that higher ranked players, in general, have more kills, which is to be expected as killing enemy heroes brings big advantages in game. The best players also tend to assist more in kills performed by others, which means that teamwork is needed and used to actually get a kill on the enemy heroes. One can notice that data is discrete, which is a consequence of average kill/assists/death statistics being available at integer values, rounded from closest real value.
  \begin{figure}[b]
\centering
  \includegraphics[width=0.48\textwidth, clip, trim=0mm 0mm 0mm 9mm]{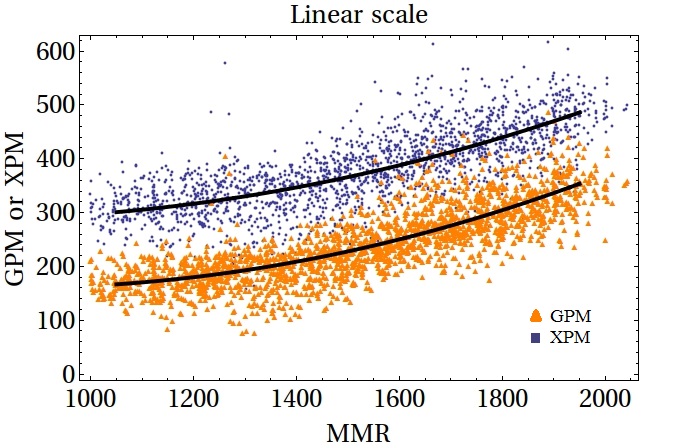}
	\caption{GPM  and XPM  as a function of MMR} 
	\label{fig:G}
		 \end{figure}   
 \subsection{Gold  and Experience}	 
The two main resources in game are gold and experience. Gold per Minute (GPM) and Experience per Minute (XPM) are metrics which describe the average amount of gold per minute and experience per minute, respectively, that a player manages to accumulate during a game. It is calculated as all the gold (experience) the player got in all the games divided with the total length of his games.
 \\
We observe a strong non-linear correlation between both GPM and XPM and skill of the players in Figure~\ref{fig:G}. One of the key ``ingredients'' of the game is the knowledge of how to maximize use of the resources, and how to increase gold and experience earned in order to strengthen the  one's character.
This is especially true for higher levels of play, where knowing the other elements of the game is not sufficient to advance if the resource management is not optimized. 

Although both GPM and XPM curves exhibit similar behaviour, ratio between them is not constant and it is constantly increasing as we go to the higher levels of MMR (Figure~\ref{fig:G1}). 		 
The reason lies in the difference between the game mechanics that determine how these resources are collected.  While to get experience one has to stand in the proximity of the enemy unit killed, to gain gold one has to actually place the killing blow on the enemy unit (known as last-hitting).
As mentioned, that does not apply exactly to killing enemy heroes, where all heroes that deal some damage in a short time frame get a small bounty. Less skilled players often miss the opportunity, or even do not realize the importance of last-hitting enemy units (who instead die to friendly creeps, defensive towers or other players), thereby only gaining the experience but not the gold (leading to observed effect).\\
\begin{figure}
\centering
  \includegraphics[width=0.48\textwidth, clip, trim=0mm 0mm 0mm 9mm]{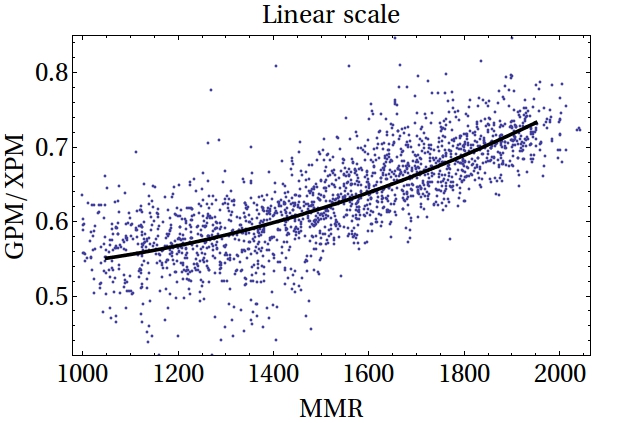}
	\caption{GPM/XPM ratio as a function of MMR} 
	\label{fig:G1}
	 \end{figure} 

\subsection{Game duration}
\label{game_dur}	 
When inspecting distribution of the game matches, it should be noted that game developers have introduced an option for one team to concede the match, but only after 15 minutes have passed. Also, conceding is easier after 30 minutes when only 4 out of 5 players need to agree to concede. This is clearly reflected in Figure~\ref{fig:A2}, which represents the game duration for the last six games played by the players from the sample. Spikes at around 15 and 30 minutes game durations are evident in the distribution. After 30 minutes, the probability for longer matches gradually declines (as games naturally tend to finish). 
\begin{figure}
\centering
  \includegraphics[width=0.48\textwidth, clip, trim=0mm 0mm 0mm 9mm]{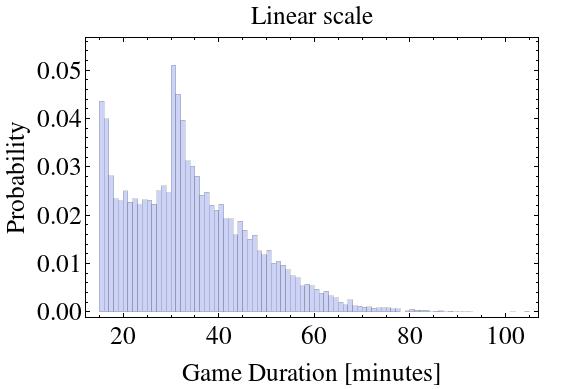}
	\caption{Distribution of games duration  } 
		\label{fig:A2}
	 \end{figure}  
\\	 		 
The average duration of the game is also clearly correlated with the skill level of players: with the higher MMR, the games tend to be shorter. We believe that is effect of higher ranked players knowing how to improve their character faster and how to utilize their advantages in the game, e.g., by destroying enemy fortifications, bringing the game to a faster conclusion. Another factor that may contribute is that higher ranked, more-experienced players can sooner realize that the game is lost and communicate this, thus reducing the likelihood of ``griefing'' (i.e., intentional harassment of other players \cite{Chesney:Griefing}) by refusing to concede.

\begin{figure}[b]
\centering
  \includegraphics[width=0.48\textwidth, clip, trim=0mm 0mm 0mm 9mm]{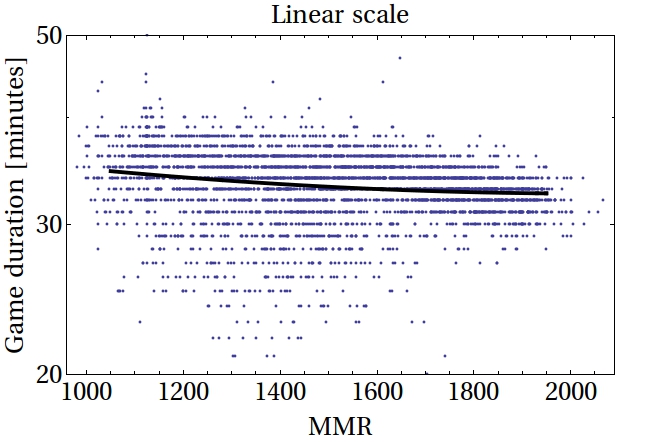}
	\caption{Average game length as function of MMR  } 
	\label{fig:GL}
	 \end{figure}

\subsection{Action rate} 
	 \label{sec:apm} 
	 	\begin{figure}
\centering
  \includegraphics[width=0.48\textwidth, clip, trim=0mm 0mm 0mm 9mm]{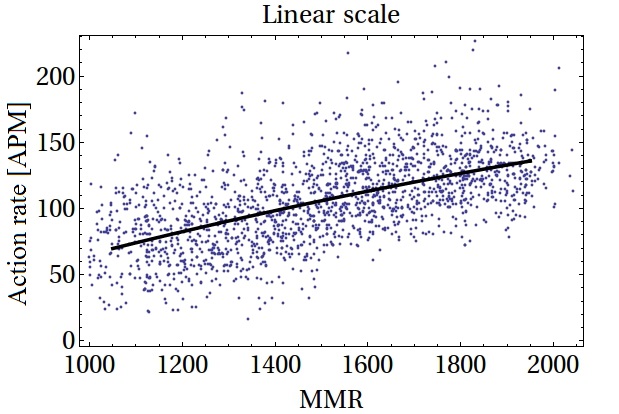}
	\caption{Action rate as a function of MMR } 
	\label{fig:A}
	 \end{figure}
The action rate in online games is commonly measured  through actions per minute (APM). Actions are clicks that player makes by using a mouse or a keyboard while in game and playing a map (changing settings or actions outside the game window are not counted). Again, this quantity is strongly correlated with the skill of the player. While lower ranked players have to think more, high end players are typically very familiar with the game and use a lot of actions to get slight positional advantage such as introducing element of randomness. Such randomness in one's position by constantly shifting slightly around disorientates enemy and makes it harder for enemy players to predict one's position.
	 
\subsection{Warding rate} 
	 	 
Wards are items that players can buy, and place in various positions on the map to give themselves and their team vision of the area around it, supplying the team with important information, such as the movements of enemy heroes. 
\\
A ward is used for benefit of the team and not for improving a single character. Additionally, a player who buys and places a ward is exposed to additional risk when venturing to place the ward. This being so, one or two players have to sacrifice themselves for the team, often composed of players unknown to them.  Also, one should note that even though the price of wards is relatively low, the game engine puts a hard cap on the amount of wards that can be bought at single time, so no team can not have a full control of the map at any single time.
\\
\begin{figure}[b]
\centering
  \includegraphics[width=0.48\textwidth, clip, trim=0mm 0mm 0mm 9mm]{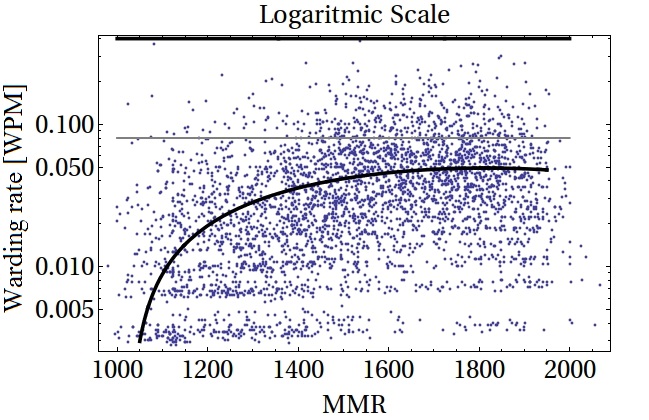}
	\caption{Warding rate as function of MMR} 
	\label{fig:W}
	 \end{figure}   
Figure~\ref{fig:W} shows the warding rate in wards placed per minute (WPM) as a function of MMR. Note that Figure~\ref{fig:W} uses the logarithmic scale. The upper thick black line is set at 0.4 wards per minute, which is the maximum number of wards per minute one can set in a typical game that last 35 minutes. The gray line is set at 0.08 wards per minute, which is 0.4 wards per minute divided by 5. This number shows the value that a ``perfect player'' would have, the player who buys wards every fifth match. Gaps that can be seen in data points are due to rounding the average number of wards per match to the closest 1/10 of the integer value. 
Figure~\ref{fig:W} shows that higher ranked players, in general, ward more. The MMR values have a large spread, which is caused by different players specializing in playing different heroes, that are more or less suitable for warding. Also, there is slight decline in the number of wards per minute for the very best players, illustrating the common practice that lower ranked players often buy wards to free the resources and time for presumably better players, who are more capable of using them for the benefit of the team. To reiterate, we observe that very good players ($\gtrsim$ 1800) when playing with even better players in their team are ready, in order to increase their chance to win a game, to take on ``less-attractive'' roles in the game (e.g., warding support) so as to free the presumably better players (ones with higher rating) to play heroes that need a lot of resources to be become effective.
\subsection{Creep denying} 
\begin{figure}
\centering
  \includegraphics[width=0.48\textwidth, clip, trim=0mm 0mm 0mm 10.5mm]{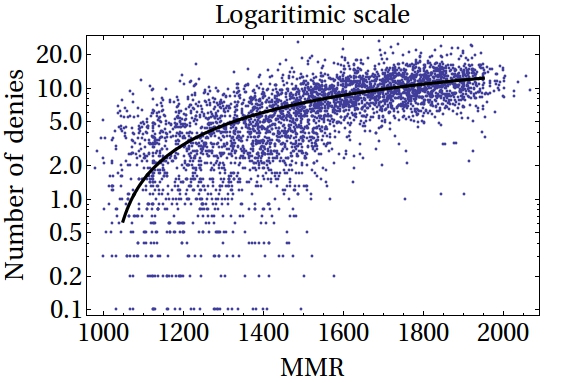}
	\caption{Number of denies in game as a function of MMR} 
	\label{fig:D}
	 \end{figure}  
Creep denying is an interesting practice in the game,
which is realized when the players destroy their own creeps, thereby ``denying'' experience and gold to the players of opposing team. Particularly useful at the start of the match, if successful, denying can lead to a slight early advantage which can be built upon to provide an even bigger mid-game advantage. In later stages, it is rarely used, as the amount of gold and experience denied becomes negligible compared to gold and experienced pilled up on every character. \\
We can observe how rating and average number of denies is strongly correlated in Figure~\ref{fig:D}. Low ranking players deny rarely, while high ranking players deny often. This is also because, at least partially, this kind of game mechanics is not immediately obvious to new players. A rising trend with higher MMR is constant throughout the whole range. Also, even if it the difference between 1800 and 1600 brackets is only in few creeps, this can really make or break a game and higher ranked players realize the importance of each creep. There are almost no high end players that do not deny.

\subsection{Account age}
We have studied the correlation of the number of games played per account and its MMR in section~\ref{game_dur} and we are interested if the results of that analysis will be similar to the correlation between the account age and MMR. Account age is defined as a time passed since the account's creation, as of the date when data were collected (November 29, 2012). Similar studies regarding the experience of chess players and their rating has been presented in \cite{vandermaaswagenmakers:PsychometricAnalysis, grabnersternneubauer:Psyhometric}.

	\begin{figure}
\centering
  \includegraphics[width=0.4\textwidth, clip, trim=0mm 0mm 0mm 10.5mm]{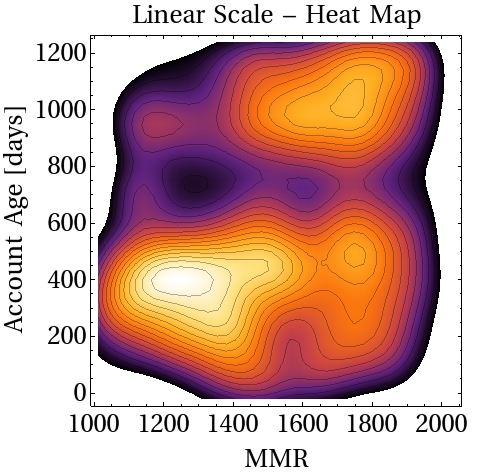}
	\caption{Heat map of the account age and MMR correlation }
		\label{fig:MT}
	 \end{figure} 
Figure~\ref{fig:MT} uses a ``heat map'' which shows the higher concentration of points with brighter colors. Also, the regions in which the value is very low have been cut off, for clarity. 
There are several interesting features to notice, as they are directly related to different payment models used during the history of the game. 
First, we notice two very distinct regions in the top and lower parts of the figure. 
HoN was in beta stage until May 12, 2010 which is 932 days before the date that data was gathered. After that, the users had to pay to create new account (pay-to-play model), until July 29, 2011 (489 days before data was gathered), when the game went to a free-to-play model (users do not pay for creating an account). During the pay-to-play phase there were fewer new accounts created, corresponding to a ``valley'' on the map, at approximately between 900 and 600 days ago. We also notice that people, who have created an account at the very start of the game history, are now, in general, in the higher bracket, while newer players are more concentrated in the lower regions. It may also be noticed that around the 400 days mark there is a surge of new accounts after the payment model changed to free to play. Also, notice there are very few of players with either a very high, or a very low ranking, who have created an account recently ($< 50$ days), indicating that it's not usual to move far from the starting rating (MMR of 1500) in a short amount of time. This is same effect as observed in Figure~\ref{fig:GP}.

\subsection{Consistency of sample}

In this section we show consistency of our sample. For demonstration purpose, we run our algorithm to create 5 different $\gtrsim $ 3000 samples and test APM/MMR correlation described in section \ref{sec:apm}. Lines for each sample are drawn in five colors (red, orange, black, green and yellow), together with the black curve used in the sample used throughout the paper. Figure~\ref{fig:C} shows that there is no significant difference between the samples.  The lines are closely aligned, and the black line is ``hiding'' most of the other lines. 
	\begin{figure}
\centering
  \includegraphics[width=0.48\textwidth]{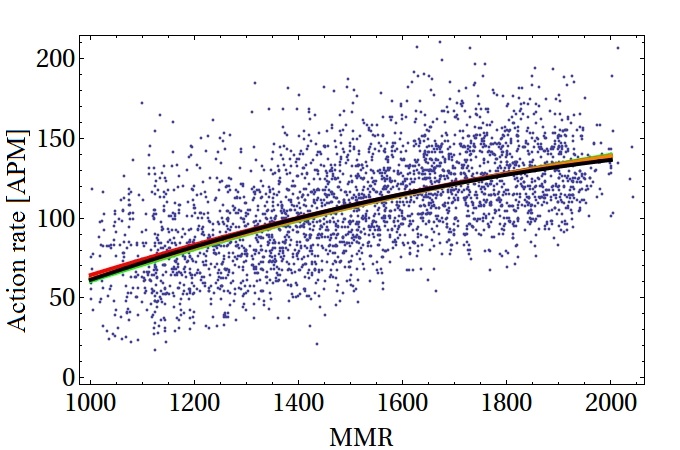}
	\caption{Action rate as a function of MMR for different samples} 
	\label{fig:C}
	 \end{figure}  

\section{Conclusion}
In this work we have evaluated the ranking system of HoN through correlating the in-game performance metrics with the rating assigned by the game's matchmaking system. Our findings indicate that the MMR system does capture the skill of the players well, but that there are some anomalies and drawbacks of the system which could be improved.

We notice some exploits which are compromising the ranking system with respect to the ``true skill'' of players, such as intentional lowering of the MMR with carefully planned groups, which is revealed by inspecting the kill/death ratio and win/loss ratio. 
Also, we find that the employed algorithm is still rather slow in placing players into particular skill groups. Therefore, we conclude that the problem of smurfs is still not solved with the current rating system.

In future work we aim to look further into player behaviour patterns and session characteristics.  
\section{Acknowledgments}
This work has been supported by the European Community Seventh Framework Programme under grant No. 285939 (ACROSS), and by the MZOS research project 036-0362027-1639 ``Content Delivery and Mobility of Users and Services in New Generation Networks''.

%
%
%
%

\balancecolumns
\end{document}